\documentstyle[prl,aps,multicol,epsf]{revtex}
\begin{document}
\draft
\widetext
\title{Ferromagnetic transition in a 
double-exchange system}
\author{Mark Auslender$^1$ and Eugene Kogan$^2$}
\address{$^1$ Department of Electrical and Computer Engineering,
Ben-Gurion University of the Negev,
P.O.B. 653, Beer-Sheva, 84105 Israel\\
$^2$ Jack and Pearl Resnick Institute 
of Advanced Technology,
Department of Physics, Bar-Ilan University, Ramat-Gan 52900, 
Israel}
\date{\today}
\maketitle
\begin{abstract}
\leftskip 54.8pt
\rightskip 54.8pt
We study ferromagnetic transition in three-dimensional double-exchange model.
The influence of strong spin fluctuations on conduction electrons is described
in coherent potential approximation. In the framework of thermodynamic approach
we construct for the  system "electrons (in a disordered
spin configuration) + spins"  the Landau functional, from the  analysis of which 
critical temperature of ferromagnetic transition is calculated. 
\end{abstract}
\pacs{ PACS numbers: 75.10.Hk, 75.30.Mb, 75.30.Vn}
\begin{multicols}{2}
\narrowtext

\section{Introduction}

The recent rediscovery of colossal magnetoresistance (CMR) in doped Mn oxides
with perovskite structure R$_{1-x}$D$_x$MnO$_3$ (R is a rare-earth metal and D
is a divalent metal, typically Ba, Sr or Ca) \cite{helmolt} 
substantially increased  interest  in the
double-exchange (DE) model \cite{zener,anderson}. Several approaches were 
used lately to  study the
thermodynamic properties of the DE model, including 
the Dynamical
Mean Field
Approximation  (DMFA) (see 
\cite{furukawa2} and references therein), Schwinger bosons \cite{arovas}, and
Variational mean-field approach \cite{alonso}. Calculated  critical
temperature $T_c$ of the ferromagnetic (FM)
transition in three-dimensional DE model,  surprisingly
well
agrees with the results of Monte Carlo method \cite{motome,alonso}, and high
temperature expansion \cite{roder} for all values of the itinerant electrons
concentration.
Still we decided to present  our 
calculation of the $T_c$. We believe that our derivation is 
interesting by itself; it also opens the opportunity
for the  analysis of the
concurrent action of the impurity potential and magnetic disorder
\cite{auskog,auskog2}.

\section{Hamiltonian and CPA equations}

We consider the DE model with the inclusion of the 
single-site impurity potential. We apply the quasiclassical 
adiabatic approximation and consider each 
core spin as a static vector
of fixed length $S$ (${\bf S}_i=S{\bf n}_i$, where ${\bf n}_i$
is a  unit vector).
The Hamiltonian  of the model $H([{\bf n}_i])$ in site representation is
\begin{eqnarray}
\label{ham}
\hat{H}_{ij}=t_{i-j}+ 
\left(\epsilon_i -J{\bf n}_i\cdot\hat{\bf \sigma}\right)\delta_{ij},
\end{eqnarray}
where $t_{i-j}$ is the electron hopping, $\epsilon_i$ is the 
on-site energy, $J$ is the effective 
exchange 
coupling between a  core spin and a conduction electron and 
$\hat{\bf \sigma}$ is the vector of the Pauli matrices. The hat above the
operator reminds that  it is a 
$2\times 2$ matrix in the spin space (we discard the
hat when the operator is a scalar matrix in the spin space).  
The Hamiltonian (\ref{ham}) is random due to randomness of 
a  core spin configuration $[{\bf n}_j]$; 
we also take into account 
the randomness of the on-site energies
$\epsilon_i$. 

To treat  Hamiltonian (\ref{ham}) we use the
coherent potential Approximation (CPA) \cite{ziman,soven,kubo,auskog}, i.e.
we  present it as
\begin{eqnarray}
\hat{H}=\overbrace{\left(t_{i-j}+\hat{\Sigma}\delta_{ij}\right)}^{H^0}+
\overbrace{\left(\epsilon_i -J{\bf n}_i\cdot\hat{\bf \sigma}-\hat{\Sigma}
\right)\delta_{ij}}^{\hat{V}}
\end{eqnarray}
(the site independent self-energy $\hat{\Sigma}(E)$ is to be determined later),
and construct a perturbation theory with respect to random potential 
$\hat{V}$.
 To do this let us
introduce  the $T$-matrix as the solution of the equation
\begin{equation}
\hat{T}=\hat{V}+\hat{V}\hat{G}_0\hat{T},
\end{equation}
where $\hat{G}_0=(E-\hat{H}^0)^{-1}$.
For the exact Green function $\hat{G}$ we get
\begin{equation}
\label{green}
\hat{G}=\hat{G}_0+\hat{G}_0\hat{T}\hat{G}_0.
\end{equation}
The coherent potential approximation (CPA)
is expressed by the equation
\begin{equation}
\label{g}
\left\langle\hat{G}\right\rangle=\hat{G}_0\Longleftrightarrow
\left\langle\hat{T}_i\right\rangle=0,
\end{equation}
where $\hat{T}_i$ is the solution of the equation
\begin{equation}
\hat{T}_i=\hat{V}_i+\hat{V}_i
\hat{g}(E-\hat{\Sigma})\hat{T}_i,
\end{equation}
and
\begin{equation}
\label{locator}
g(E)=\left(G_0(E)\right)_{ii}=
\int\frac{N_0(\varepsilon)}{E-\varepsilon}d\varepsilon,
\end{equation} 
where $N_0(\varepsilon)$ is the bare density of states.
Equation for the $T$-matrix can be presented as an algebraic equation for the
$2\times 2$ matrix 
$\hat{\Sigma}$:
\begin{equation}
\label{gencpa}
\left\langle\left[1-\hat{V}_i\hat{g}(E-\hat{\Sigma})\right]^{-1}
\hat{V}_i\right\rangle_{{\bf n},\epsilon}=0.
\end{equation}
Eq. (\ref{gencpa}) 
coincides exactly with that of Ref. \cite{furukawa}.

\section{Landau functional of a double-exchange system}

In this part  we start from 
the exact partition function of the system electrons + spins 
\begin{equation}
Z=\int\exp\left[-\beta F_e([{\bf n}_j])+\beta{\bf H}\sum_j{\bf n}_i\right]D{\bf
n}_j ,
\end{equation}
where $N$ is the number of core spins, ${\bf H}$ is magnetic field,
$\beta=1/T$, and $F_e([{\bf n}_j])$ is the free energy
of the electron subsystem  
for a given core spin configuration $[{\bf n}_j]$. Our aim is to obtain Landau
functional of the system.

In the mean-field approximation all energy levels depend only upon macroscopic
magnetization ${\bf m}=\frac{1}{N}\sum_j{\bf n}_j$. 
(Mean-field  means that we do not take into account large scale  fluctuations 
of macroscopic magnetization, but we certainly take into account microscopic
fluctuations of ${\bf n}_i$.)  
Hence
\begin{equation}
F_e([{\bf n}_j])\approx F_e(m),
\end{equation}
and the partition function can be written as
\begin{equation}
\label{part}
Z=\int \exp\left[-\beta F({\bf m},{\bf H})\right]d{\bf m} ,
\end{equation}
where
\begin{equation}
\label{part2}
F({\bf m},{\bf H})=F_e(m)-N{\bf H}{\bf m}-TS(m);
\end{equation}
the quantity
\begin{equation}
\label{eent}
S(m)=\ln\int 
\delta\left({\bf m}-\frac{1}{N}\sum_j{\bf n}_j\right)D{\bf n}_j
\end{equation}
is the entropy of the core spin subsystem.
We can identify the exponent with the
Landau functional of the whole system \cite{negele}. Further on we'll be interested in the case
${\bf H}=0$; in this case $F({\bf m},{\bf H})=F(m)$.

At high temperatures the minimum of the functional is at $m=0$. At low
temperatures 
the point $m=0$ corresponds to the maximum
of the functional. So if we expand $F_e$ and $S$ with respect to $m$
\begin{eqnarray}
F_e(m)=F_0-F_2m^2+\dots\nonumber\\
S(m)=S_0-S_2m^2+\dots,
\end{eqnarray}
the critical temperature $T_c$ is found from the equation
\begin{equation}
\label{tc}
F_2=T_cS_2.
\end{equation}
In fact, we need only the quadratic terms of the Landau functional to find the
$T_c$.

Here it is appropriate to note that in the  approach  of Ref.
\cite{furukawa}
the thermodynamic part of the solution is based on the
equation
\begin{equation}
\left\langle{\bf n}\right\rangle=\int d\Omega_{\bf n}P({\bf n}){\bf n},
\end{equation}
where $P({\bf n})$ is the Boltzmann factor for a spin, determined from the DMFA
ansatz. 

To calculate the entropy  we  expand the 
$\delta$-function in Eq. (\ref{eent}) in Fourier integral
\begin{eqnarray}
\delta\left({\bf m}-\frac{1}{N}\sum_j{\bf n}_j\right)\sim
\int e^{i\left(N{\bf m}-\sum_j{\bf n}_j\right){\bf r}}d{\bf r},
\end{eqnarray}
which gives an
opportunity to calculate integrals with respect to $D{\bf n}_j$. After simple algebra
we obtain  (ignoring
irrelevant multipliers in the argument of the logarithm)
\begin{eqnarray}
S(m)=\ln \left\{(4\pi)^N\int_0^{\infty} \sin(Nrm)
\left(\frac{\sin r}{r}\right)^Nrdr\right\}.
\end{eqnarray}
Calculated integral by the steepest descent method, up to the
terms of order of $m^2$ we obtain
\begin{eqnarray}
\label{classic}
S=N\left[\ln(4\pi)-\frac{3m^2}{2}\right].
\end{eqnarray}

The grand canonical potential for the electron subsystem is
\begin{equation}
-\beta \Omega(\mu, m)=\frac{N}{\pi}\int_{-\infty}^{+\infty}\ln
\left[1+e^{-\beta(E-\mu)}\right]\mbox{Im}\;g_c(E)dE,
\end{equation}
where  $g_c$ is the charge 
locator $g_c=\mbox{Tr}\;\hat{g}$.
The free energy is
\begin{equation}
F_e(N,m)=\Omega(\mu, m)+\mu N;
\end{equation}
the connection between the  number of holes per cite $x$ 
and chemical potential $\mu$ is given by the equation
\begin{equation}
\label{fermi}
1-x=\frac{1}{\pi}\int_{-\infty}^{\infty}f(E)\mbox{Im}\;g_c(E)dE,
\end{equation}
where $f(E)$ is the Fermi distribution
function.

The calculation is especially simple, when we can consider the electron gas as
degenerate and ignore the terms of order of $T/\mu$ (that is we can consider the
temperature of electron gas as being effectively equal to zero). In this case
\begin{equation}
F_e=\frac{N}{\pi}\int_{-\infty}^{\mu}E\;\mbox{Im}g_c(E)dE.
\end{equation}
Expanding  $g_c$ with respect to $m$ 
\begin{equation}
\label{exp}
g_c=g_0+g_2m^2+\dots,
\end{equation}
we obtain
\begin{equation}
\label{f2}
F_2=\frac{N}{\pi}\int_{-\infty}^{\mu_0}(\mu_0-E)\;\mbox{Im}g_2(E)dE,
\end{equation}
where $\mu_0$ is found from the equation
\begin{equation}
\label{fermi2}
1-x=\frac{1}{\pi}\int_{-\infty}^{\mu_0}\mbox{Im}\;g_0(E)dE.
\end{equation}

\section{$T_c$ for semi-circular  density of states}

We consider semi-circular (SC) bare (i.e. for $\epsilon\equiv 0$ and $J=0$) DOS 
\begin{equation}
N_{0}( \varepsilon) =\frac{2}{\pi W}
\sqrt{1-\left( \frac{\varepsilon}{W}\right) ^{2}},
\label{scdos}
\end{equation}
at $\left| \varepsilon\right| \leq W$ and $N_{0}(\varepsilon) =0$ 
otherwise, for
which 
\begin{equation}
\label{gf}
g(E) =\frac{2}{W}\left[\frac{E}{W}-
\sqrt{\left( \frac{E}{W}\right)^{2}-1}\right].  
\label{scgrfun}
\end{equation}
Hence we obtain
\begin{equation}
\label{sigma}
\hat{\Sigma}=E-2w\hat{g}-\hat{g}^{-1},
\end{equation}
where $w=W^2/8$.
Thus, Eq. (\ref{gencpa}) can be presented as 
\begin{equation}
\label{basic}
\hat{g}=\left\langle\left(E-\epsilon+J{\bf n}{\bf \sigma}
-2w{\hat g}\right)^{-1}
\right\rangle_{{\bf n},\epsilon}.
\end{equation}
It is convenient to write the locator $\hat{g}$ in the form
\begin{equation}
\hat{g}=\frac{1}{2}(g_c\hat{I}+{\bf g}_s\hat{{\bf \sigma}}),
\end{equation}
where $\hat{I}$ is a unity matrix. For the charge locator $g_c$ and spin locator
 ${\bf g}_s$ we
obtain the system of equations
\begin{eqnarray}
\label{basic2}
g_c=2\left\langle\frac{E-2wg_c}{\left(E-\epsilon-wg_c\right)^2-
(J{\bf n}+w{\bf g}_s)^2}
\right\rangle_{{\bf n},\epsilon}\nonumber\\
{\bf g}_s=2\left\langle\frac{J{\bf n}+w{\bf g}_s}{\left(E-\epsilon-wg_c\right)^2-
(J{\bf n}+w{\bf g}_s)^2}
\right\rangle_{{\bf n},\epsilon}.
\end{eqnarray}

In the strong Hund coupling limit 
($J\rightarrow \infty $) 
we obtain from Eqs. (\ref{basic2}) two decoupled spin sub-bands. For
each sub-band, after shifting the energy by $\pm J$ we obtain
\begin{eqnarray}
\label{system}
g_c=\left\langle\frac{1}
{E-\epsilon-wg_c-wn^zg_s}\right\rangle_{{\bf n},\epsilon}\nonumber\\
g_s=\left\langle\frac{n^z}
{E-\epsilon-wg_c-wn^zg_s}\right\rangle_{{\bf n},\epsilon},
\end{eqnarray}
where  ${\bf g}_s=(0,0,g_s)$ (axis OZ is directed along ${\bf m}$). 

For a saturated FM phase ($m=1$), we obtain $g_s=g_c$, and 
closed equation for the charge locator  is
\begin{eqnarray}
g_c=\left\langle\frac{1}
{E-\epsilon-2wg}\right\rangle_{\epsilon}.
\end{eqnarray}
For a paramagnetic (PM) phase ($m=0$), we obtain ${\bf g}_s=0$, and 
closed equation for the charge locator $g_c(m=0)\equiv g_0$ is
\begin{equation}
\label{infJcpa}
g_0=\left\langle\frac{1}
{E-\epsilon-wg_0}\right\rangle_{\epsilon}.
\end{equation}
In both cases only the averaging with respect to random on-sight energies is 
left. So, the DOS in PM phase  is equal to the DOS in
FM phase of  another model, with decreased by a factor of 
$\sqrt{2}$ bare bandwidth.

From here on we ignore the on-site
energies disorder and put $\epsilon\equiv 0$. In this case 
the solution of Eq. (\ref{infJcpa}) for
$g_0$  coincides with Eq. (\ref{gf}), with $W/\sqrt{2}$ substituting for $W$.
Expanding Eq. (\ref{system}) 
with respect to $m$ we obtain the system
\begin{eqnarray}
g_2m^2=g_0^2(wg_2+mwg_s)+g_0^3\frac{w^2}{3}g_s^2,\nonumber\\
g_s=mg_0+g_0^2\frac{w}{3}g_s.
\end{eqnarray}
Substituting the value of  $g_2$ into Eq. (\ref{f2}), 
we obtain the critical temperature:
\begin{equation}
\label{ttc}
T_c=\frac{2w}{3\pi}\int_{-\infty}^{\mu_0}(\mu_0-E)\mbox{Im}
\left[\frac{g_0^3}
{(1-wg_0^2) \left(1-\frac{w}{3}g_0^2\right)^2}\right]dE.	
\end{equation}
Calculated integral analytically, we get for the $T_c$:
\begin{eqnarray}
\label{tcw}
T_{c} =\frac{\sqrt{2}}{4\pi }\left[ \sqrt{1-y^{2}}-\frac{1}{\sqrt{3}}\tan^{-1} 
\sqrt{3(1-y^{2})}\right]W 
\end{eqnarray}
where the reduced chemical potential
$y=\sqrt{2}\mu_0/W$
is found from the equation
\begin{equation}
x=\frac{1}{2}-\frac{1}{\pi}\left[\sin^{-1}y+y\sqrt{1-y^2}\right].
\end{equation}
\begin{figure}
\epsfxsize=2.5truein
\centerline{\epsffile{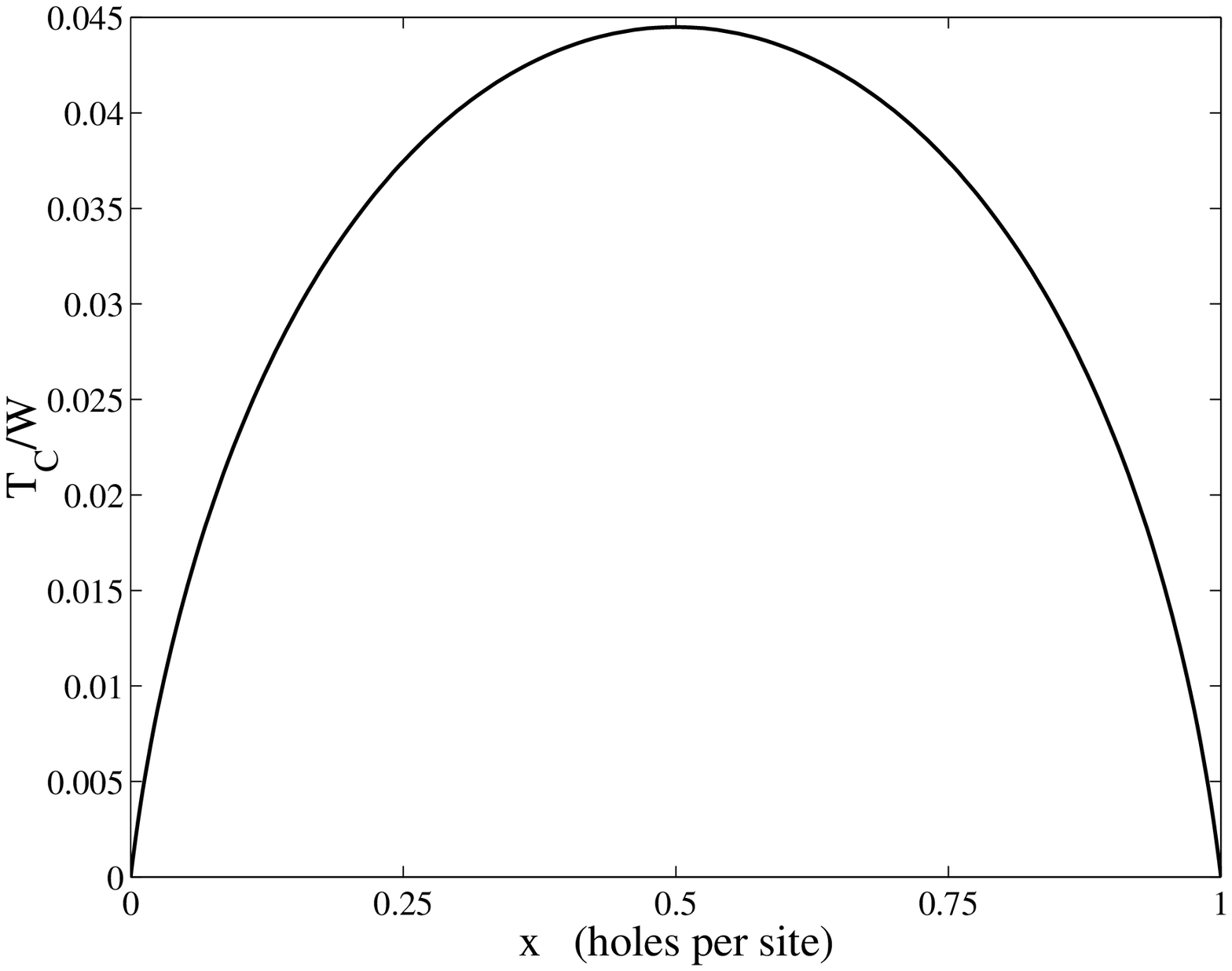}}
\label{Fig.1}
\end{figure}
\noindent
{\footnotesize {\bf Fig. 1.} 
The calculated Curie temperature for the semi-circular DOS at $J=\infty$ as
function of hole concentration $x$.} 

One notices the particle-hole symmetry, namely $T_c(x)=T_c(1-x)$. At $x\to 0$
and $x\to 1$, $T_c\to 0$. The optimal doping  turns out to be $0.5$,
with $T_c=0.045W$.

\section{Discussion}

Any calculation of the $T_c$  in the DE model includes, on
one hand, calculation of the energy of  electron  interacting with
randomly directed core spins,  and, on the other hand, the analysis
 of phase transition in the system of
spins with the interaction inferred from the solution of the electron part of the
problem.  The difference
between our approach and that of Ref. \cite{furukawa}, is similar to the
difference between    
the two  mean-field approximations  for the Heisenberg model: 
one based on the  analysis of a local spin in a coherent field, which  by itself depends
upon the averaged magnetization, the other based on the
expansion of the partition function \cite{ziman,negele}. 
For the Heisenberg model the two approaches give exactly the same result. 
This turned out to be the case
also the case for the DE model. 
Our result for the $T_c$ (Eq. (\ref{ttc})), using integration by parts, can be exactly reduced to Eq. (49)
 of Ref. \cite{furukawa}, which in
our notation is
\begin{equation}
T_c=\frac{2w}{3\pi}\int_{-\infty}^{\mu_0}\mbox{Im}
\left[\frac{g_0^2}
{1-\frac{w}{3}g_0^2}\right]dE.	
\end{equation}
The result of Refs.
\cite{arovas} for the $T_c$ in our notation is
\begin{equation}
T_c=\frac{\sqrt{2}}{9\pi}\left(1-y^2\right)^{3/2}W. 
\end{equation} 
It is close to our result numerically, but the functional dependence of the
$T_c$ upon hole
concentration 
is essentially different. The difference comes from
our use of the CPA for the description of electrons, in distinction from  the
mean-field decoupling of charge and core spin fluctuations in de
Gennes' spirit, used in Refs. \cite{arovas}. (In the part of  
Refs. \cite{arovas} where the charge-spin coupling is taken into account, the
explicit result for the $T_c$ is not presented.)

The  DE scenario of the FM transition in manganites 
attracted especial attention
after its validity was questioned in Ref. \cite{millis}. In the following
discussion,  the position of one side can be expressed by reformulation of
the rhetorical 
question by Hubbard \cite{hubbard}:
"How, in the  itinerant model, explain a Curie temperature $\sim 1000^\circ$ for
iron, when calculations always give an exchange field $\sim 1-2$ eV?" in the 
form:
How, in the DE model, explain a Curie temperature $\sim 300^\circ$ for
a manganite (at optimal doping), when calculations always give a bandwidth 
$\sim 1-2$ eV? The fallacy is based on the implicit assumption that the $T_c$
should be of the order of the energy difference between the  fully ordered FM
state and the paramagnetic state (which is the smallest of the band width
and the exchange integral). It does not take into account that
the fluctuations which
restore magnetic symmetry in itinerant models  are local fluctuations 
in the orientation of magnetic
moments, with   much lower  energy \cite{hubbard}.
That is why the numerical coefficient in Eq. (\ref{tcw}) turns out to be much
less than one.

\section{Acknowledgment}

We are thankful for helpful comments to J.L. Alonso, L.A. Fernandez, G.
Gomes-Santos, F. Guinea,
V. Laliena and V. Martin-Mayor. 
This research was supported by the Israeli Science Foundation administered
by the Israel Academy of Sciences and Humanities.

\end{multicols}
\end{document}